
\def\pd{\partial}
\def\ts{\thinspace}
\magnification=1200
\pageno=0
\parskip 3 pt plus 1pt minus 1 pt
\rightline{DTP-92/35}
\rightline{July 1992}
\vskip 2 true cm
\centerline{FINITE EULER HIERARCHIES AND INTEGRABLE UNIVERSAL EQUATIONS}
\vskip 2.5 true cm
\centerline{J. GOVAERTS%
\footnote{$\sp {\dag}$}
{\rm{Address {}from 1st October 1992:}\hfill\break
\it{Institut de Physique Nucl\'eaire, Universit\'e Catholique de
Louvain,\hfill\break
B-1348 Louvain-la-Neuve (Belgium)}}}
\vskip 0.5 true cm
\centerline{\it{Department of Mathematical Sciences}}
\centerline{\it{University of Durham, Durham DH1 3LE, UK}}
\vskip 2 true cm
\centerline{Abstract}
\vskip 1 true cm
Recent work on Euler hierarchies of field theory Lagrangians iteratively
constructed {}from their successive equations of motion is briefly reviewed.
On the one hand, a certain triality structure is described, relating
arbitrary field theories, {\it classical\ts} topological field theories
-- whose classical solutions span topological classes of manifolds -- and
reparametrisation invariant theories  -- generalising ordinary string
and membrane theories. On the other hand, {\it finite} Euler hierarchies are
constructed for all three classes of theories. These hierarchies terminate
with {\it universal\ts} equations of motion, probably defining new integrable
systems as they admit an infinity of Lagrangians. Speculations as to the
possible relevance of these theories to quantum gravity are also suggested.
\vskip 1.5 true cm
\centerline{To appear in the Proceedings of the Colloquium}
\centerline{``Quantum Groups"}
\centerline{Charles University, Prague, 18 - 20 June 1992}
\vfill\eject
\vskip 10pt
\leftline{\bf 1. Introduction}
\vskip 10 pt
Using techniques of triangulations of Riemann surfaces and integrations over
random matrices, recent developments in string theory have exposed$\sp {[1]}$
profound connections between two dimensional quantum gravity coupled to a
variety of matter fields, two dimensional topological gravity and classically
integrable systems in two dimensions. On the other hand, it was the problem of
quantising the latter classes of theories that also brought$\sp {[2]}$ quantum
groups and related algebraic structures into the arena of present day
mathematical physics. Motivated by these results, some time ago we
considered$\sp {[3]}$ the two dimensional Bateman equation$\sp {[4]}$ for a
field $\phi(x,y)$ function of variables $x$ and $y$:
$$\phi_{xx}\ts\phi_y\sp 2 + \phi_{yy}\ts\phi_x\sp 2
- 2 \phi_{xy}\ts\phi_{x}\ts\phi_{y} = 0\ .\eqno(1.1)$$
This nonlinear equation possesses interesting properties, which are easily
established {}from its determinant form
$$\det\pmatrix{0&\phi_x&\phi_y\cr
           \phi_x&\phi_{xx}&\phi_{xy}\cr
           \phi_y&\phi_{xy}&\phi_{yy}\cr} = 0\ .\eqno(1.2)$$
This makes it obvious that the equation is covariant under arbitrary
$GL(2)$ linear transformations of the base space coordinates $x$ and $y$,
showing that such transformations map solutions into one another. More
importantly however, the equation is also covariant under redefinitions of
the field as $\phi\rightarrow\varphi=F(\phi)$, with $F(\phi)$ a totally
arbitrary function. This property is one of general covariance in the
target space parametrised by $\phi$, which shall thus be referred to as such.
Hence, under field redefinitions, solutions to (1.1) fall into topological
classes of the target space. The Bateman equation is a simple example of a
{\it classical topological field theory\ts}$\sp {[5]}$, namely a field theory
whose space of {\it classical} solutions spans topological
classes of manifolds. Classical topological field theories
may be viewed as being halfway between ordinary field
theories and {\it quantum} topological field theories$\sp {[6]}$.

There is still another property of the Bateman equation reminiscent of
quantum topological field theories. Namely, (1.1) follows$\sp {[3]}$
{}from the massless Klein-Gordon equation in $2+1$ dimensions
$$\phi_{tt}=\phi_{xx}+\phi_{yy}\ ,\eqno(1.3a)$$
by imposing the nonlinear constraint
$$\phi_t\sp 2=\phi_x\sp 2+\phi_y\sp 2\ .\eqno(1.3b)$$
However, these are the equations of motion for the action
$$\int{dt\ts dx\ts dy\ {1\over 2}(1+\lambda)
(\phi_t\sp 2-\phi_x\sp 2-\phi_y\sp 2)}\ ,\eqno(1.4)$$
where $\lambda$ is a Lagrange multiplier for the constraint $(1.3b)$. This
action has the peculiarity of vanishing identically on the constraint surface,
a property quite similar to that of topological field theories whose actions
are typically given by surface terms$\sp {[7]}$. Actually, when represented
as in (1.3), the Bateman equation is also reminiscent of the Nambu-Goto
string$\sp {[8]}$. Indeed, in the conformal gauge, the string equations of
motion are
$$\phi_{\tau\tau}\sp \mu=\phi_{\sigma\sigma}\sp \mu\ ,\eqno(1.4a)$$
whereas the constraints of reparametrisation invariance in the world-sheet are
$$\phi_{\tau}\sp 2+\phi_{\sigma}\sp 2=0=\phi_{\tau}\phi_{\sigma}\eqno(1.4b)$$
($\phi\sp \mu$ are the Minkowski spacetime string coordinates while $\tau$ and
$\sigma$ are the usual world-sheet variables). Except for
the second condition in $(1.4b)$, (1.4) is in direct correspondence with
(1.3). Finally, given that the Nambu-Goto action for strings and membranes
reads
$${\cal L}_{\rm{Nambu-Goto}}
=\bigl[\det {}_{(ij)} N\sp t N\bigr]\sp {1\over 2}\ ,\eqno(1.5)$$
where $N\sp \mu{}_j=\pd\phi\sp \mu/\pd x_j$, examples of classical
topological field
theories generalising the Bateman equation are simply obtained {}from the
action
$${\cal L}=\bigl[\det {}_{(\mu\nu)} N N\sp t\bigr]\sp {1\over 2}
\ ,\eqno(1.6)$$
where the product of $N\sp \mu{}_j$ with its transposed is taken in
the reversed
order. The equations of motion of (1.6) indeed transform covariantly under
field redefinitions
$\phi\sp \mu\rightarrow\varphi\sp \mu=F\sp \mu(\phi\sp \nu)$ with
$F\sp \mu$ being arbitrary functions of all fields. In fact, as will be
discussed
later on, reparametrisation invariant field theories, of which Nambu-Goto
actions are simple examples, and classical topological field theories of the
type considered in this paper, of which (1.6) are also simple examples,
are theories dual$\sp {[5]}$ to one another.

Integrability  of the Bateman equation is made explicit by considering the
variable $u=\phi_x/\phi_y$, in terms  of  which  the  equation reduces to
$$u_x=u\ts u_y\ .\eqno(1.7)$$
Note how this choice of variable also  makes  the  general  covariance  of
(1.1) under arbitrary field redefinitions manifest, since (1.7) only involves
the  field  $u(x,y)$ which   is   invariant   under  field  redefinitions  of
$\phi(x,y)$. Integrability (\`a la Liouville) of the Bateman equation is
also  manifest  {}from  (1.7). Indeed,  this  equation is a
simple reduction of the KdV equation in which the term linear in $u_{yyy}$
is absent. The system thus possesses an infinity of conserved currents and
charges,  namely  for every integer  $n$
$$\pd_x  u\sp n = \pd_y\bigl[{n\over{n+1}}\  u\sp {n+1}\bigr]\ ,\eqno(1.8)$$
which are in involution for either of the  two  symplectic structures  of
the  KdV  hierarchy  (the  second  structure thus also lacking the central
extension term). In fact, all solutions to the Bateman equation may be defined
implicitly, either by the constraint
$$x\ts F(\phi) + y\ts G(\phi) = {\rm constant}\ ,\eqno(1.9)$$
in the $\phi$-representation ($F(\phi)$, $G(\phi)$ and ``constant" being
arbitrary), or by the constraint
$$u = H(x u + y)\ ,\eqno(1.10)$$
in the $u$-representation ($H(t)$ being arbitrary).

Finally, the Bateman equation also admits a variational formulation through an
action principle. In fact, there exists an {\it infinity \ts} of inequivalent
(i.e. not differing by surface terms) local Lagrangian densities all leading
to (1.1) as their equation of motion! This is quite remarkable indeed, since
in the generic case only systems with a {\it single} degree of freedom admit
an infinity of Lagrangian formulations$\sp {[9]}$, whereas we are dealing here
with a field theory! The Lagrangian densities whose equation of motion is the
Bateman equation are {\it all} weight one homogeneous functions of the first
derivatives $\phi_x$ and $\phi_y$, namely
${\cal L}(\lambda\phi_x,\lambda\phi_y)=\lambda\ {\cal L}(\phi_x,\phi_y)$.
Note that even though (1.1) is covariant under $GL(2)$ transformations and
field redefinitions, these transformations need not be symmetries of any of
these Lagrangians (${\cal L}$ always scales with a factor $F'(\phi)$ under
field redefinitions $\phi\rightarrow F(\phi)\ts$). The Bateman equation is
thus an example of a system whose space of classical solutions admits more
symmetries than its Lagrangian. In fact, the universality of the Bateman
equation associated to this infinity of Lagrangian formulations is also at
the origin of its integrability. In the general case of an arbitrary
number of fields $\phi\sp a(x_i)\ (a=1,\cdots,D;\ i=1,\cdots,d\ts)$ and
variables
$x_i$, the Euler-Lagrange equations of motion are given by the Euler operators
$${\cal E}_a {\cal L} = -{\pd{\cal L}\over\pd\phi\sp a}
+ \pd_i\bigl[{\pd{\cal L}\over\pd\phi\sp a_i}
        -\pd_j{\pd{\cal L}\over\pd\phi\sp a_{ij}} + \cdots\bigr]
\ .\eqno(1.11)$$
Hence, for any Lagrangian whose dependence on the fields is only through
their derivatives but not explicitly on the fields themselves -- only
Lagrangians of this type are considered in this work --, the equations of
motion are simply conservation equations for a collection of currents. When
the same equations admit an infinity of inequivalent Lagrangians, we then
have in fact an infinity of currents and charges which are conserved for
classical solutions (integrability \`a la Liouville still requires involution
of all these charges for any sympletic structure(s) associated to
Hamiltonian formulations of the same system. There are presumably an infinity
of such structures for systems admitting an infinity of Lagrangians).
In the case of the Bateman equation, this remark explains why all integer
powers of $u$ define conserved charge densities. Indeed, we may always
choose ${\cal L}(\phi_x,\phi_y) = \phi_y\ F(\phi_x/\phi_y)$
with $F(u)$ arbitrary, so that
$${\pd{\cal L}\over\pd\phi_x}=F'({\phi_x\over\phi_y})\quad,
\ {\pd{\cal L}\over\pd\phi_y}=F({\phi_x\over\phi_y})
- {\phi_x\over\phi_y}\ F'({\phi_x\over\phi_y})\ .\eqno(1.12)$$
Thinking now in terms of a power series expansion of $F(\phi_x/\phi_y)$ in
the variable $u=\phi_x/\phi_y$, and using (1.12) in (1.11), one recovers
indeed the conservation equation (1.8).

We have succeeded$\sp {[3,10,5]}$ in generalising the above attractive
properties of the Bateman equation to situations for which the numbers of
fields and variables are arbitrary. In doing so, we also discovered some
interesting new structures -- of which the Bateman equation is but the
simplest illustration -- described as follows.
On the one hand, there exist transformations between arbitrary
field theories (whose Lagrangians do not explicitly depend on the fields but
only on their derivatives), reparametrisation invariant field theories and
classical topological field theories. These transformations define a closed
triality diagram relating all three types of theories and their classical
solutions in a unique manner. These are the C- and R-maps and their inverse
maps briefly described below. On the other hand,
specific Euler hierarchies of field theories also play a distinguished
r$\hat{\rm o}$le.
Euler hierarchies$\sp {[3,5]}$ are hierarchies of Lagrangian field theories in
which each theory is constructed in an arbitrary way out of the equations of
motion of the theory at the previous level in the hierarchy. Namely, there
exist$\sp {[3]}$ {\it finite} Euler hierarchies, {\it i.e.} hierarchies
terminating after a {\it finite} number of iterations of the procedure just
described, and ending with equations of motion which are {\it universal},
{\it i.e.} which are {\it independent} of the arbitrariness inherent to the
construction of the Euler hierarchy.
Thus by construction, these equations admit an {\it infinity}
of inequivalent Lagrangians. Moreover, since all Lagrangians of these finite
Euler hierarchies depend on derivatives of fields only, for the reasons
discussed above, this raises the strong possibility that these universal
equations are actually new integrable systems in arbitrary dimensions.
Finally, using these results and the aformentioned triality, one has a way of
constructing dual finite Euler hierarchies of arbitrary field theories, of
classical topological field theories and of reparametrisation invariant ones,
all leading to universal equations of motion. In particular, in this manner
one obtains new string and membrane theories with dynamics being described
by universal equations of motion which are most certainly integrable, in
contradistinction to Nambu-Goto membranes.

In this short note, it is not possible to present all the considerations
relating to these results, for which the reader is referred back to the
original work$\sp {[3,10,5]}$. Here, we shall merely sketch the main
conclusions. First, the close analogy between properties of Lagrangians for
reparametrisation invariant field theories and for classical topological ones
is presented in the next section. In the following two sections,
the C- and R-maps are briefly described. In sect.5, the generic finite Euler
hierarchy for one field is constructed, leading through the C- and R-maps to
the associated two dual hierarchies, whose universal equations of motion are
respectively a generalisation of the Bateman equation to an arbitrary number
of dependent variables and a universal $(d-1)$-dimensional membrane in
$(d+1)$ dimensions. To conclude, we briefly comment on further
results$\sp {[3,5]}$ concerning arbitrary numbers of fields, while also
suggesting directions for further investigations.
\vskip 15pt
\leftline{\bf 2.  Homogeneity, Reparametrisation Invariance
and General Covariance}
\vskip 10pt
Generally, consider theories of $D\ts$ fields $\phi\sp a$ depending on
$d\ts$ coordinates $x_i$, with Lagrangian densities
${\cal L}(\phi\sp a_i,\phi\sp a_{ij})$ functions of first and second
derivatives of the fields only. Even though the results of this section
remain valid for Lagrangians depending on derivatives of arbitrary
higher order, the restriction to first and second derivatives only is
sufficient for our purposes. The equations of motion for the fields
$\phi\sp a$ are thus given by ${\cal E}_a{\cal L}[\phi\sp a]=0$, where
the Euler operators ${\cal E}_a$ are given in (1.11).

First, let us consider Lagrangians with the following homogeneity property
$$
{\cal L}(R_i{}\sp j \phi\sp a_j,R_i{}\sp k R_j{}\sp l \phi\sp a_{kl}
+T\sp k_{ij}\phi\sp a_k) =
(\det R_i{}\sp j)\sp \alpha
{\cal L}(\phi\sp a_i,\phi\sp a_{ij})\ .\eqno(2.1)$$
Here, $R_i{}\sp j$ and $T\sp k_{ij}$ are arbitrary coefficients,
and $\alpha$ is the weight of homogeneity of the Lagrangian.
Using this property and the identities following {}from it by
differentiation with respect to the parameters $R_i{}\sp j$ and
$T\sp k_{ij}$, a straightforward calculation shows that the equations
of motion obey the following relations
$$\phi\sp a_i{\cal E}_a{\cal L}[\phi\sp a]=(\alpha-1)\ \pd_i{\cal L}
(\phi\sp a_i,\phi\sp a_{ij})\ .\eqno(2.2)$$
Therefore, when $\alpha=1$ there are $d\ts$ identities among the $D\ts$
equations of motion, leaving only $(D-d\ts)$ independent equations of motion.
In fact, the case $\alpha=1$ corresponds to a reparametrisation invariant
action. Any Lagrangian obeying (2.1) scales under reparametrisations with
a factor which for $\alpha=1$ precisely cancels the reparametrisation
Jacobian for the integration measure $\prod_i dx_i$ over the coordinates.
Hence in this case, the identities
$$\phi\sp a_i{\cal E}_a{\cal L}[\phi\sp a]=0\ ,\eqno(2.3)$$
are the Noether (or Ward) identities$\sp {[11]}$ due to reparametrisation
invariance, leaving only $(D-d\ts)$ independent equations of motion.

By analogy with (2.1), consider Lagrangians with the following homogeneity
property
$${\cal L}(\phi\sp b_iR_b{}\sp a,\phi\sp b_{ij}R_b{}\sp a
   +\phi\sp b_iT\sp {ba}_j+\phi\sp b_jT\sp {ba}_i)=
(\det R_a{}\sp b)\sp \alpha {\cal L}(\phi\sp a_i,\phi\sp a_{ij})\ .\eqno(2.4)$$
Here, $R_a{}\sp b$ and $T\sp {ab}_i$ are arbitrary coefficients, and
$\alpha$ is the
weight of homogeneity of the Lagrangian. Given this property, consider now
the transformation of the equations of motion ${\cal E}_a{\cal L}[\phi\sp a]$
under arbitrary field redefinitions
$\phi\sp a\rightarrow\varphi\sp a=F\sp a(\phi\sp b)$.
Using the homogeneity property (2.4) and the identities following {}from it
by differentiation with respect to $R_a{}\sp b$, $T_i\sp {ab}$,
$\phi\sp a_i$ and
$\phi\sp a_{ij}$, a straightforward calculation then shows that
the equations of
motion ${\cal E}_a{\cal L}[\varphi\sp a]$ for the transformed
fields $\varphi\sp a$
are given in terms of those for the fields $\phi\sp a$ by
$${\cal E}_a{\cal L}[\varphi\sp a]=
(\det R_a{}\sp b)\sp \alpha (R\sp {-1})_a{}\sp b
{\cal E}_b{\cal L}[\phi\sp a]
+(R\sp {-1})_a{}\sp b{\pd(\det R_a{}\sp b)\sp \alpha
\over\pd\phi\sp b}(\alpha-1)
{\cal L}(\phi\sp a_i,\phi\sp a_{ij})\ ,\eqno(2.5)$$
where $R_a{}\sp b=\pd F\sp b/\pd\phi\sp a$.
Therefore, whenever the Lagrangian possesses the homogeneity property (2.4)
with a weight $\alpha=0$ or $\alpha=1$, the equations of motion for the
fields $\phi\sp a$ transform covariantly among themselves under arbitrary
field redefinitions. In particular in the case of one field $D=1$, the
equation of motion is {\it invariant} when $\alpha=1$. Hence, any Lagrangian
${\cal L}(\phi\sp a_i,\phi\sp a_{ij})$ homogeneous in the sense of (2.4)
with a weight $\alpha=1$ or $\alpha=0$ defines a classical topological field
theory. In fact, the case $\alpha=1$ plays a distinguished r$\hat{\rm o}$le,
as will become clear in the next section when discussing the C-map.
\vskip 15pt
\leftline{\bf 3. The C-Map}
\vskip 10pt
To define the C-map, let us consider an arbitrary field theory Lagrangian
${\cal L}(y\sp a_\alpha,y\sp a_{\alpha\beta})$ dependent on the first and
second derivatives of $p\ts$ fields $y\sp a(x_\alpha)$ functions of $q\ts$
coordinates $x_\alpha$. The associated action is thus
$$S[y\sp a]=\int{\prod_\alpha dx_\alpha \ {\cal L}
({\pd y\sp a\over\pd x_\alpha},{\pd\sp 2y\sp a\over\pd x_\alpha\pd x_\beta})}
\ .\eqno(3.1)$$
Let us now introduce $p\ts$ additional coordinates $\phi\sp a$, and extend
the coordinate dependence of the fields $y\sp a$ to also include a
dependence on these new variables, {\it i.e.} $y\sp a(x_\alpha,\phi\sp b)$.
Nevertheless, the
corresponding field theory is still described by the original Lagrangian
${\cal L}(y\sp a_\alpha,y\sp a_{\alpha\beta})$, with the action now given
as in (3.1) with however an additional integration over the variables
$\phi\sp a$. In particular, the equations of motion for this new theory
are {\it identical} to those of (3.1) since the Lagrangian is independent
of any derivatives of the fields $y\sp a$ with respect to the new
variables $\phi\sp a$. In other words, as far as the dynamical evolution
of the fields $y\sp a$ is concerned, the variables $\phi\sp a$ are irrelevant.

The C-map is implemented by inverting the $\phi\sp a$-dependence of the
fields $y\sp a$, and by considering $\phi\sp a$ as functions of
$x_i=(x_\alpha,y\sp a)$,
{\it i.e.} $\phi\sp a(x_i)$. This requires that the matrix of derivatives
${\pd y\sp a}/{\pd\phi\sp b}$ be non singular, or equivalently that we have
$\det M_a{}\sp b\neq 0$ with $M_a{}\sp b=\pd\phi\sp b/\pd y\sp a$.
By direct substitution in the Lagrangian for the original field theory
$y\sp a(x_\alpha,\phi\sp b)$ and its action,
one then obtains a Lagrangian for a theory of $D=p\ts$ fields
$\phi\sp a(x_i)$ dependent on $d=p+q\ts$ independent variables
$x_i=(x_\alpha,y\sp a)$. In fact, as may be expected, the resulting theory
is a classical topological field theory since, as was pointed out above,
the $\phi\sp a$-dependence in the original theory is irrelevant, so that the
equations of motion for the fields $\phi\sp a$ in the new theory should be
independent of the parametrisation used for these fields, namely their
equations of motion should be generally covariant under arbitrary field
transformations. Indeed, it is easy to check$\sp {[5]}$ that the Lagrangian
obtained through the C-map obeys the homogeneity property (2.4) with a weight
$\alpha=1$.

The C-map thus takes an arbitrary field theory of $p\ts$ fields in $q\ts$
dimensions into a classical topological field theory of $D=p\ts$ fields in
$d=p+q\ts$ dimensions. However, if the original theory is itself already a
classical topological field theory obeying (2.4) with $\alpha=1$,
the C-map only reproduces$\sp {[5]}$ this
original theory, in agreement with the fact that the inversion defining the
C-map is essentially a redefinition of the fields $y\sp a$. Moreover, it is
also possible$\sp {[5]}$ to define an inversion of the C-map, namely a
transformation taking any classical topological field theory obeying (2.4)
with $\alpha=1$ and $D=p,\ d=p+q\ts$ into an arbitrary field theory
with $D=p\ts$ and $d=q\ts$. In turn, the image of this latter theory under
the C-map is again the original classical topological field theory. Hence,
the C-map establishes a one-to-one correspondence between arbitrary
field theories and classical topological field theories obeying (2.4)
with $\alpha=1$ and having fewer fields than coordinates.
\vskip 15pt
\leftline{\bf 4. The R-Map}
\vskip 10pt
To define the R-map, consider now an arbitrary field theory Lagrangian
${\cal L}({\pd\phi\sp \mu\over\pd z_i},
{\pd\sp 2\phi\sp \mu\over\pd z_i\pd z_j})$
dependent on the first and second derivatives of $p\ts$ fields
$\phi\sp \mu(z_i)$
functions of $q\ts$ coordinates $z_i$. The associated action is thus
$$S[\phi\sp \mu]=\int{\prod_i dz_i \ {\cal L}
({\pd\phi\sp \mu\over\pd z_i},{\pd\sp 2\phi\sp \mu\over\pd z_i\pd z_j})}\ .
\eqno(4.1)$$
Let us now introduce $q\ts$ arbitrary functions $x_i(z_j)$, and extend the
field content of the theory to also include these extra degrees of freedom
while still keeping the same Lagrangian and action as in (4.1). Obviously, the
equations of motion for $\phi\sp \mu$ remain as before, whereas those for the
new fields $x_i$ are trivially satisfied since the Lagrangian is independent
of these degrees of freedom.

Nevertheless, by inverting the $z_i$ dependence of the theory, one obtains a
new field theory with reparametrisation invariance. For this purpose,
consider field configurations such that
${\det M_i{}\sp j}\neq 0$ with $M_i{}\sp j=\pd z_j/\pd x_i$.
The $z_i$ dependence of the fields $x_i$ may then be inverted, leading to a
field theory of $D=p+q$ fields $\phi\sp a=(\phi\sp \mu,z_i)$ functions of
$d=q$ variables $x_i$, with an action obtained {}from (4.1) by direct
substitution. The resulting field theory is reparametrisation invariant in
the coordinates $x_i$. Clearly, this is to be expected since on the one hand,
the choice for the functions $x_i(z_j)$ is totally arbitrary, and on the
other hand, their equations of motion are trivial so that the transformed
theory should indeed be independent of the choice of parametrisation in
$x_i$. {}From a geometric point of view, the functions $\phi\sp \mu(z_i)$
define a $q\ts$-dimensional subspace of the $(p+q)$-dimensional space
spanned by the coordinates $\phi\sp a=(\phi\sp \mu,z_i)$. Introducing
the fields $x_i(z_j)$ amounts to introducing an arbitrary intrinsic
parametrisation of this subspace, without affecting its topological and
geometrical properties as an embedded space. In other words, we are
simply dealing with parametrised $(q-1)$-dimensional membrane
theories in $(p+q)$ dimensions.

The series of operations described above define the R-map, which
thus takes any field theory of $p\ts$ fields in $q\ts$ dimensions into
a reparametrisation invariant field theory of $D=p+q\ts$ fields in $d=q\ts$
dimensions. Again, it is straightforward to verify$\sp {[5]}$ that the
transformed Lagrangian obtained {}from (4.1) indeed satifies the homogeneity
property (2.1) with weight $\alpha=1$. On the other hand, as is the case
for the C-map, applying the R-map on a theory which is already
reparametrisation invariant only reproduces$\sp {[5]}$ the latter, the
inversion defining the R-map being indeed a reparametrisation in the
coordinates $z_i$. Similarly, it is also possible to the define$\sp {[5]}$
the inverse R-map taking
any reparametrisation invariant theory of $D=p+q\ts$ fields in $d=q\ts$
dimensions into an arbitrary field theory with $D=p\ts,\ d=q\ts$, whose
image under the R-map is again the original reparametrisation invariant
theory. Hence, the R-map puts arbitrary field theories and reparametrisation
invariant ones in one-to-one correspondence.

Composition of the C- and R-maps is obviously also possible. The following
equivalences have already been mentioned:
$$C\circ C\ \sim\ C,\ \quad R\circ R\ \sim\ R\ .\eqno(4.2)$$
Moreover, the successive application of the two maps leads$\sp {[5]}$ to the
further equivalences
$$C\circ R\ \sim\ C,\ \quad R\circ C\ \sim\ R\ .\eqno(4.3)$$
Therefore, we have indeed the triality structure described in the
introduction. Namely, given an arbitrary field theory (whose Lagrangian
only depends on derivatives of fields), there exists a triplet of dual
theories which is closed under the action of the C- and R-maps and their
inverses. In addition to the
original theory, the C-map produces a classical topological field theory
obeying (2.4) with $\alpha=1$ and fewer fields than coordinates, whereas the
R-map produces a reparametrisation invariant field theory with more fields
than coordinates (whose number of independent equations is that of the
original theory). Moreover, the C- and R-maps put the latter two theories
in one-to-one correspondence.
\vskip 15pt
\leftline{\bf 5. The Generic Finite Hierarchy}
\vskip 10 pt
Consider a collection $F_n(\phi_i)\ (n=1,2,\ldots)$ of arbitrary functions
of the first derivatives of a field $\phi(x_i)$ dependent on $d\ts$
coordinates $x_i$. The associated Euler hierarchy of Lagrangians is defined
recursively by
$${\cal L}_n=F_n\ W_{n-1},\qquad W_0=1\ ,\eqno(5.1)$$
where the $W_n\ (n=1,2,\ldots)$ are the equations of motion
$$W_n={\cal E}{\cal L}_n\ ,\eqno(5.2)$$
with the Euler operator ${\cal E}$ defined in (1.11).

The fundamental result is the following identity$\sp {[5,3]}$:
$$W_n=
\ {1\over(d-n)!}\ \epsilon_{i_1\cdots i_d}\ \epsilon_{j_1\cdots j_d}
M\sp {(1)}_{i_1 k_1}\cdots M\sp {(n)}_{i_n k_n}
\ \phi_{k_1 j_1}\cdots\phi_{k_n j_n}
\ \delta_{i_{n+1}j_{n+1}}\cdots\delta_{i_d j_d}\ ,\eqno(5.3)$$
where
$$M\sp {(p)}_{ij}=\ {\pd\sp 2 F_p\over\pd\phi_i\pd\phi_j}\ ,
\quad p=1,2,\cdots,n\ .\eqno(5.4)$$

Consequently, $W_n$ is symmetric in the arguments
$M\sp {(p)}\ (p=1,2,\cdots,n\ts)$.  Namely, the order in which the
multiplicative factors $F_n(\phi_i)$ are introduced in the hierarchy is
irrelevant. Moreover, the dependence of $W_n$ on these functions and on
the second derivatives $\phi_{ij}$ separates$\sp {[5]}$. In particular,
at level $\ell=d\ts$ we obtain
$$W_d=\ \epsilon_{i_1\cdots i_d}\ \epsilon_{j_1\cdots j_d}
M\sp {(1)}_{i_1 j_1}\cdots M\sp {(d)}_{i_d j_d}\ {\det \phi_{ij}}
\ ,\eqno(5.5)$$
showing that the dependence on the functions
$F_n(\phi_i)\ (n=1,2,\cdots,d\ts)$
factorizes at this level, leading to the following {\it universal} equation
of motion for the Lagrangian
${\cal L}_d$ independently of the factors $F_n(\phi_i)\ (n=1,2,\cdots,d\ts)$
$${\det \phi_{ij}}=0\ .\eqno(5.6)$$
Since the choice for the functions $F_n(\phi_i)\ (n=1,2,\cdots,d\ts)$ is
totally arbitrary, (5.6) is indeed an example of an equation of motion
admitting an infinite number of inequivalent Lagrangians. Moreover, since
$W_d$ is always a surface term -- being the equation of motion for a
Lagrangian without an explicit dependence on the field $\phi$ --, one also
concludes that there is an infinite number of inequivalent conserved
currents for the universal equation (5.6), suggesting its possible
integrability. Indeed for $d=2$, (5.6) is a particular reduction of
Plebanski's equation$\sp {[12]}$ for self-dual gravity in four dimensions,
a system known to be integrable.

The result (5.6) also implies that the hierarchy terminates at that level.
Indeed, given any Lagrangian of the form
$${\cal L}(\phi_i,\phi_{ij})=F(\phi_i)\ {\det \phi_{ij}}\ ,\eqno(5.7)$$
as is ${\cal L}_{d+1}$, it is easily shown that its equation of
motion ${\cal E}{\cal L}$ vanishes identically. Hence, the universal equation
(5.6) is also the last non trivial equation of motion for the hierarchy.
The recursive construction in (5.1) and (5.2) always leads to
$W_n=0$ for $n\geq{d+1}$. This concludes the construction of the generic
finite Euler hierarchy leading to the universal equation (5.6).

Using the C- and R-maps, it is now possible to construct two more dual
finite Euler hierarchies also leading to universal equations. The C-map leads
to a hierarchy of classical topological field theories for one field all
obeying (2.4) with $\alpha=1$. The R-map leads to a hierarchy of
reparametrisation invariant field theories with one more field than
independent coordinates, namely string and membrane theories. In fact, the
former hierarchy is constructed as in (5.1) and (5.2) with the only further
restriction that the
functions $F_n(\phi_i)$ are now homogeneous and of weight $\alpha=1$
but are otherwise arbitrary. Consequently, all equations of motion
$W_n$ are invariant under redefinitions of the field $\phi$ (see sect.2).
However, this hierarchy of classical topological field theories terminates
at level $\ell = d-1$ rather than $\ell = d$ as is the case for the generic
hierarchy, with nevertheless the universal equation of motion
$${\det\pmatrix{0&\phi_j\cr\phi_i&\phi_{ij}\cr}}=0\ ,\eqno(5.8)$$
up to an overall factor encapsulating all the dependence on the functions
$F_n(\phi_i)$ much as in (5.5). Clearly, (5.8) is a generalisation of the
Bateman equation (1.1) to higher dimensions, the latter equation corresponding
to $d=2$ and thus to a hierarchy terminating at level $\ell = 1$ for any
choice of weight one homogeneous function as initial Lagrangian, in agreement
with the claims of the introduction.

The R-map applied to the generic hierarchy (5.1) produces reparametrisation
invariant theories of $D=d+1$ fields $\phi\sp a(x_i)$ of $d\ts$ variables
$x_i$. It is convenient to define the Jacobians
$$J_a=\ (-1)\sp d\ \epsilon_{a b_1\cdots b_d}
\ \phi_1\sp {b_1}\cdots \phi_d\sp {b_d}\ .\eqno(5.9)$$
Consider then a collection $F_n(\phi\sp a_i)\ (n=1,2,\ldots)$ of arbitrary
functions with the homogeneity property
$$F_n(R_i{}\sp j\phi_j\sp a)=(\det R_i{}\sp j)\ F_n(\phi_i\sp a)
\ .\eqno(5.10)$$
The associated finite Euler hierarchy is then obtained {}from the iterative
procedure
$${\cal L}_n=\ F_n \ W_{n-1}\ ,\quad W_0 = 1\ ,\eqno(5.11)$$
where, due to reparametrisation invariance, the equations of motion always
factorize$\sp {[10,5]}$ with the Jacobians $J_a$ so that for any fixed $a$,
here not summed over,
$$W_n=\ {1\over J_a}\ {\cal E}_a{\cal L}_n\ .\eqno(5.12)$$
This hierarchy of reparametrisation invariant field theories thus terminates
at level $\ell = d\ts$ with an equation which, up to an overall factor
containing all the dependence on the functions $F_n$, factorizes into the
universal equation of motion (implicit summation over $a$ is again
understood)
$$\det {}_{(ij)}(\phi\sp a_{ij} J_a)=0\ .\eqno(5.13)$$
In fact, due to the Noether identities (2.3), there is only one
independent equation of motion at each level of the hierarchy (essentially
$W_n$ in (5.12)\ts), which at level $\ell = d\ts$ leads to (5.13).
The universal equation of this hierarchy thus describes the dynamics of new
$(d-1)$-dimensional membranes in a $(d+1)$-dimensional space which are most
probably integrable systems.
\vskip 15pt
\leftline{\bf 6. Concluding Remarks}
\vskip 10pt
It is possible to extend the previous results to other finite Euler
hierarchies leading to universal equations of motion in the case of
arbitrary theories with more than one field. In turn, through the
C- and R-maps, one obtains new
finite Euler hierarchies of classical topological and reparametrisation
invariant field theories, with universal equations, for which the difference
between the numbers of fields and coordinates is arbitrary.
The idea$\sp {[5]}$ is simply to consider the generic hierarchy with the
single field $\phi$ being the
sum of all fields $\phi\sp a$, {\it i.e.} $\phi=\phi\sp a\lambda\sp a$,
and then expand all equations in the parameters $\lambda\sp a$. This leads
to further generalisations of the Bateman equation in terms of classical
topological field theories, as well as their dual reparametrisation
invariant counterparts.

However, this does not answer the question whether there might still exist
other independent finite Euler hierarchies with universal equations, for
example with Grassmann odd degrees of freedom or supersymmetrisations of the
ones above. Another possibility might be the extension of the Euler operator
to some kind of noncovariant exterior derivative on field space, since for
Lagrangians independent of the fields themselves the Euler operator is
actually nilpotent$\sp {[3,13]}$, {\it i.e.} ${\cal E}\sp 2=0$,
a property essential for the existence of the
generic hierarchy above. On the other hand, would all field theories
admitting an infinity of Lagrangians be the universal equations of some
universal finite Euler hierarchy?

The whole issue of integrability of the equations constructed here remains to
be solved. This question has been answered$\sp {[5]}$ in the affirmative in
the simplest cases, but otherwise only general classes of solutions have been
obtained$\sp {[5]}$ so far. Obviously, this problem is of interest since the
present equations would probably define new integrable systems in arbitrary
dimensions and as is well known, it is the study of classical and quantum
integrability of two dimensional systems that led to the development of new
mathematical techniques and the discovery of new algebraic structures such as
quantum groups$\sp {[2]}$.

Finally, motivated by more physical considerations,
the problem of quantising the equations constructed here
is also of interest. On the one hand for classical topological field
theories, an understanding of their quantum states
could turn out to be of relevance to quantum gravity.
As opposed to quantum topological field theories where it
is hoped$\sp {[6]}$ that the geometrical structure of spacetime would result
{}from some phase transition, in the context of classical topological field
theories one may envisage the possibility that spacetime geometry is actually
a {\it quantum} phenomenon appearing only once these theories of spacetime
topology are quantised! On the other hand, since ordinary critical strings
always describe at least gravitational interactions, the BRST
quantisation$\sp {[11]}$ and physical spectrum of the new universal string
and membrane equations appearing in our work is certainly a problem
worth investigating as well, having in mind again possible applications to
theories of gravity.
\vskip 15pt
\leftline{\bf Acknowledgements}
\vskip 10pt
The author would like to thank A. Morozov and especially D. B. Fairlie for a
very enjoyable and stimulating collaboration, and
the organisers of the Colloquium for their efficient organisation of a most
pleasant and informative meeting. This work is supported through a Senior
Research Assitant position funded by the S.E.R.C.
\vfill\eject
\leftline{\bf REFERENCES}
\vskip 10pt
\frenchspacing
\item{[1]}For a recent discussion, see\hfil\break
Witten E.: Surveys in Diff. Geom. {\it 1} (1991) 243.
\item{[2]}See for example\hfil\break
Faddeev L., Reshetikhin N., Takhtajan L.: {\it in} Braid Groups, Knot Theory
and Statistical Mechanics,
Advanced Series in Mathematical Physics {\it 9} (Eds. C. N. Yang and
M. L. Ge). World Scientific, Singapore, 1989, p. 97.
\item{[3]}Fairlie D. B., Govaerts J., Morozov A.: Nucl. Phys.
{\it B373} (1992) 214.
\item{[4]}Bateman H.: Proc. Roy. Soc. London {\it A125} (1929) 598.
\item{[5]}Fairlie D. B., Govaerts J.: Durham preprint DTP-92/17 (April 1992),
hepth/9204074. To appear in Jour. Math. Phys. (1992).
\item{[6]}Witten E.: Commun. Math. Phys. {\it 117} (1988) 353.
\item{[7]}Labastida J. M. F., Pernici M.: Phys. Lett. {\it B212} (1988) 56;
Labastida J. M., Pernici M., Witten, E.: Nucl. Phys. {\it B310} (1988) 611.
\item{[8]}Nambu Y.: Lectures at the Copenhagen Symposium (1970);
Goto T.: Prog. Theor. Physics {\it 46} (1971) 1560.
\item{[9]}Morandi G., Ferrario C., Lo Vecchio G., Marmo G., Rubano C.:
Physics Reports {\it 188} (1990) 147.
\item{[10]}Fairlie D. B., Govaerts J.: Phys. Lett. {\it B281} (1992) 49.
\item{[11]}For a recent review, see \hfil\break
Govaerts J., Hamiltonian Quantisation and Constrained Dynamics,
Lecture Notes in Mathematical and Theoretical Physics {\it 4}.
Leuven University Press, Leuven, 1991.
\item{[12]}Finley J. D. III, Plebanski J. F.: Jour. Math. Phys.
{\it 17} (1976) 585.
\item{[13]}Olver P. J.: Applications of Lie Groups to Differential Equations,
Graduate Texts in Mathematics {\it 107}. Springer Verlag, Berlin, 1986,
p. 252.
\end